\documentclass[aps,12pt,showpacs]{revtex4}
\usepackage{graphicx}
\usepackage{graphics}
\usepackage{color}
\usepackage{amsfonts}

\definecolor{lightlightblue}{rgb}{.85,1,1}
\def\G{\mathcal{G}}

\def\ra{\rightarrow}

\def\be{\begin{equation}}
\def\ee{\end{equation}}

\def\bea{\begin{eqnarray}}
\def\eea{\end{eqnarray}}

\def\G{\mathcal{G}}
\def\A{\mathcal{A}}

\def\O{\mathcal{O}}

\begin{document}

\title{Two-dimensional spanning webs as (1,2) logarithmic minimal model}

\author{J.G. Brankov} \email{brankov@theor.jinr.ru} \affiliation{Bogoliubov Laboratory of Theoretical Physics, JINR,
141980 Dubna, Russia \\Institute of Mechanics, Bulgarian
Academy of Sciences, 1113 Sofia, Bulgaria}
\author{S.Y. Grigorev} \affiliation{Bogoliubov Laboratory of Theoretical Physics, JINR,
141980 Dubna, Russia}\author{V.B. Priezzhev} \affiliation{Bogoliubov Laboratory of
Theoretical Physics, JINR, 141980 Dubna, Russia}
\author{I.Y. Tipunin} \affiliation{Lebedev Physics Institute, Russian Academy
of Sciences,  117924 Moscow, Russia}

\begin{abstract}
A lattice model of critical spanning webs is considered for the finite cylinder geometry.
Due to the presence of cycles, the model is a generalization of the known spanning
tree model which belongs to the class of logarithmic theories with central charge $c=-2$.
We show that in the scaling limit the universal part of the partition function for closed
boundary conditions at both edges of the cylinder coincides with the character of symplectic
fermions with periodic boundary conditions and for open boundary at one edge and closed at
the other coincides with the character of symplectic fermions with antiperiodic boundary
conditions.

\end{abstract}
\pacs{ 05.40.-a, 02.50.Ey, 82.20.-w}

\maketitle

\noindent \emph{Keywords}: free fermion model, logarithmic conformal field theory,
spanning graphs, dense polymers, Virasoro module.

\section{Introduction}

The mathematical problem of spanning trees on a connected graph can be
considered as a model of statistical mechanics and, as such, it is the first non-trivial
exactly solved multidimensional problem thanks to the famous Kirchhoff's theorem \cite{Kirchhoff}.
In modern classification, the model belongs to the class of free-fermion models \cite{FanWu}
which admit determinant solutions.
The spanning trees are associated with a variety of models, such as the Abelian sandpile
\cite{Sandpile}, Hamiltonian walks on the Manhattan lattice and dense polymer
models \cite{Manhattan, densepolymers}. The enumeration of spanning trees on the two-dimensional
square lattice is
equivalent to the close packed dimer problem solved by Kasteleyn \cite{Kasteleyn} and Temperley
and Fisher \cite{TempFish}. In the scaling limit, correlation properties of the spanning trees can be
described by the conformal field theory with central charge $c=-2$ \cite{LCFT1}-\cite{LCFT4}.

The simplest generalization of spanning trees are the spanning webs, spanning subgraphs of a connected
graph containing cycles together with tree branches attached to them \cite{FK}.
There are two sources for the appearance of topologically different classes of
cycles in models associated
with spanning trees. First, periodic boundary conditions in at least one spatial dimension generate
cycles. Such cycles appear in the exact solutions of the dimer problem on lattices wrapped on a cylinder or
torus \cite{Kasteleyn} and they are \emph{non-contractible} to a point in the embedding surface. Second, lattice
defects like monomers in a dense dimer packing give rise to a different kind of \emph{contractible}
cycles \cite{BBGJ, PPR}. In general, the spanning web model does not belong to the free fermion class and,
moreover, it is not exactly solvable. However, for particular geometries of the cycles and appropriate
statistical weights of the configurations it retains the free-fermion properties.

In this paper we calculate the partition function of a spanning web model
on a finite cylinder by considering the number of cycles winding around the cylinder as a parameter.
Our aim is to evaluate the leading finite-size corrections to the free energy
in the limit of large perimeter
of the cylinder. In the absence of cycles the finite-size effects of the spanning tree model confirm
predictions of the logarithmic conformal field theory \cite{IPRH}. We show that the presence of cycles
changes the Casimir effect in accordance with conformal weights which appear
in the Kac table \cite{PRZ}. An example of exactly solvable logarithmic models with conformal
boundary conditions has been given recently by Pearce and Rasmussen \cite{PearceRas}. They considered
critical dense polymers with certain types of defects on a strip and reproduced the conformal
weights in the first column of the extended Kac table. Their results were obtained by means of a functional
equation for commuting
transfer matrices, formulated in terms of the planar Temperley-Lieb algebra.

Despite the similarity between dense polymers at the free fermion point and spanning trees on
an auxiliary sublattice, the classification of conformal weights in these two models is quite
different. The entries of the Kac table for the model of dense polymers are labeled by the number
of "defect" lines \cite{PearceRas} under fixed boundary conditions at both sides of the strip.
We will show that the cycles in the spanning webs play the role of pairs of defect lines in the model of
dense polymers. However, in our case the boundary conditions are different for the odd and even
entries in the first column of the Kac table. Some complication of the boundary conditions is the fee
to be paid for simplicity of derivation of the partition function of the spanning web model. Analytical
calculations in Sections II and III are reduced to the standard determinant expressions for the
free fermion model with subsequent analysis by use of the Euler-Maclaurin formula.
In Section IV we show that the partition function calculated for a finite
lattice with different boundary conditions coincides with the characters of
coinvariants calculated in different modules for the algebra of symplectic
fermions. This allows us to identify open and closed boundary conditions
with modules generated by integer and half-integer modes of fermions respectively.

\section{The spanning webs model}

We consider the labeled graph $\G =(V,E)$ with vertex set $V$ and
set of bonds $E$. The vertices are sites of the square lattice
$s_{x,y}, 1 \leq x \leq M, 1 \leq y \leq N $ from which we obtain a
graph on a cylinder by identifying $s_{x,y}$ and $s_{x+M,y}$ for all
$x,y$. The graph $\G$ represents a finite square lattice embedded in
a cylinder of height $N$ and perimeter $M$, with closed boundary
conditions at the top and bottom edges. The term 'closed' means the
absence of bonds connecting vertices of $V$ with an exterior of
$\G$. We shall consider also the
case of open boundary conditions at the vertices $B \subset V$
belonging to one of the edges $\{s_{x,1},1\leq x \leq M\}$ and
$\{s_{x,N},1\leq x \leq M\}$ of the cylinder, or to both of them.
These cases correspond
to a graph $\G'=(V',E')$ with vertex set $V'=V\cup g$ containing an
additional vertex, the root $g$, and the set of bonds $E'=E\cup
\{(j,g): j\in B\}$ enlarged with the bonds connecting the vertices
of $B$ with the root $g$. For convenience of notation we label the
boundary conditions by the superscript $(\mu,\nu)$: $\mu =0$ ($\mu
=1$) denotes closed (open) top boundary and $\nu =0$ ($\nu =1$)
closed (open) bottom boundary, respectively. We find it
convenient to construct the desired spanning web configurations on
the above graphs by using the arrow representation, see e.g.
\cite{Pr85}. Accordingly, to each vertex $i \in V$ we attach an
arrow directed along one of the bonds $(i,i')$ incident to it. Each
arrow defines a directed bond $(i\ra i')$ and each configuration of
arrows $\A$ on $\G$ defines a spanning directed graph (digraph)
$\G_{sd}(\A)$  with set of bonds $E_{sd}(\A)=\{(i\ra i'): i,i'\in
V\}$ depending on $\A$. Similarly, the arrow configurations on $\G'$
define a spanning digraph $\G'_{sd}(\A)$ with set of bonds
$E'_{sd}(\A)=\{(i\ra i'): i\in V, i'\in V\cup g\}$. Note that no
arrow is attached to vertex $g$, thus it has out-degree zero. A
cycle of length $k$ is a sequence of directed bonds
$(i_1,i_2),(i_2,i_3),(i_3,i_4), \dots, (i_k,i_1)$ where all $i_j$,
$1 \leq j \leq k$ are distinct. If both $(i\ra i')$ and $(i'\ra i)$
belong to the spanning web we say that it contains a cycle of length
2. Our aim is to study sets of spanning digraphs with no other
cycles than those which wrap the cylinder. The relevant
configurations will be enumerated with the aid of a generating
function defined as the determinant of an appropriately constructed
weight matrix.

\subsection{Cylinder with closed boundaries}

We begin with the examination of the determinant expansion of the
usual Laplace matrix $\Delta$ for the graph $\G$.
Let the vertices of the set $V$ be labeled in arbitrarily order from 1
to $n=|V|=MN$. Then $\Delta$ has the elements ($i,j \in \{1, \ldots, n\}$)
\begin{equation}
\label{laplacian}
\Delta_{ij} = \left\{ \begin{array}{rll}
z_i,& \mathrm{if}& \, i=j, \\
-1,&  \mathrm{if}& \, i\, \, \mathrm{and}\, \, j \, \, \mathrm{are}\, \,
\mathrm{adjacent},\\
0,& &\mathrm{otherwise}.
\end{array}      \right.
\end{equation}
where $z_i$ is the order of vertex $i$. Since the matrix $\Delta$
has a zero eigenvalue, its determinant vanishes. On the other
hand, the Leibniz formula expresses the determinant of $\Delta$ as
a sum over all permutations $\sigma$ of the set $\{1, 2,\dots ,
n\}$:
\begin{equation}
\det \Delta = \sum_{\sigma \in S_n} \; \mathrm{sgn} (\sigma) \:
\Delta_{1,\sigma(1)} \Delta_{2,\sigma(2)} \ldots
\Delta_{n,\sigma(n)} =0,
\label{Leibniz}
\end{equation}
where $S_n$ is the symmetric group, $\mathrm{sgn} (\sigma)=\pm 1$
is the signature of the permutation $\sigma$. The identity
permutation $\sigma = \sigma_{\mathrm{id}}$ in (\ref{Leibniz}) yields
the term $z_1 z_2
\cdots z_n$ equal to the number of all possible arrow configurations
on $\G$.

In general, each permutation $\sigma \in S_n$ can be factored into a product
(composition) of disjoint cyclic permutations, say, $\sigma = c_1
\circ c_2  \cdots \circ c_k$. This representation partitions the
set of vertices $V$ into non-empty disjoint subsets - the orbits
$\O_i$ of the corresponding cycles $c_i$, $i=1,\dots ,k$. More
precisely, if $\O_i=\{v_{i,1},v_{i,2}, \dots v_{i,l_i}\}
\subset V$ is the orbit of $c_i$, then $\cup_{i=1}^k\O_i
=V$ and $\sum_{i=1}^k l_i =n$, where $l_i$ is the
cardinality of the orbit $\O_i$, equivalently, the length
of the cycle $c_i$. The orbits consisting of just one
element, if any, constitute the set $S_{fp}(\sigma)$ of fixed
points of the permutation: $S_{fp}(\sigma) = \{v=\sigma (v), v
\in V\}$. In the case of the identity permutation $\sigma_{\mathrm{id}} \in S_n$
all orbits consist of exactly one element,
$\O_i(\sigma_{\mathrm{id}})=\{v_i\}\subset V$, $i=1, \dots ,n$, and $S_{fp}
(\sigma_{\mathrm{id}})= V$.
A cycle $c_i$ of length $|c_i|=l_i \geq 2$ will be called a
\textit{proper cycle}. A proper cycle of length 2 corresponds
to two oppositely directed edges which connect a pair of adjacent
vertices: $(v_{i,1}\ra v_{i,2})$, $(v_{i,2}\ra v_{i,1})$. Note that the vertices
of an orbit $\O_i$ of cardinality $l_i = |O_i(\sigma)| \geq 3$ are connected
by a closed path on $\G$ which can be traversed in two opposite
directions: if $c_i$ is the cycle defined by $v_{i,1} \ra \sigma(v_{i,1}) =
v_{i,2}, \ra \dots \ra \sigma(v_{i,l_i})= v_{i,1}$, then the reverse
cycle $c'_i$ can be represented as $v_{i,l_i} \ra
\sigma(v_{i,l_i}) = v_{i,l_i -1}, \ra \dots \ra \sigma(v_{i,1})=
v_{i,l_i}$.

Now we take into account that the proper cycles on $\G$ are of
even length only, hence, the signature of every permutation in the
expansion of the determinant depends on the number of proper
cycles in its factorization, i.e., if $\sigma = c_1 \circ c_2
\cdots \circ c_p$, where $|c_i|\geq 2$, $i=1,\dots , p$, then
$\mathrm{sgn} (\sigma)= (-1)^p$. Thus, the terms in Eq.
(\ref{Leibniz}) can be rearranged according to the number $p$ of
disjoint proper cycles as follows: \be
\prod_{i=1}^n z_i =  \sum_{p=1}^{[n/2]} (-1)^{p+1} \sum_{\sigma = c_1 \circ
\cdots \circ c_p} \; \prod_{i=1}^{p}
\Delta_{v_i,c_i(v_i)} \Delta_{c_i(v_i),c_i^2(v_i)}\cdots
\Delta_{c_i^{l_i -1}(v_i),v_i}\prod_{j\in S_{fp}(\sigma)}z_j.
\label{altern}
\ee
Here $c_i^k$ is the $k$-fold composition of the cyclic permutation
$c_i$ of even length $l_i$, $v_i \in \O_i(\sigma)$, so that
$c_i^{k-1}(v_i)\not= c_i^k (v_i)$ and $c_i^{l_i}(v_i) = v_i$.
Note that all non-vanishing off-diagonal elements are equal to $-1$.

The above expansion
reveals the following features: (i) As expected, all spanning digraphs on
$\G$ have at least one proper cycle; (ii) Each term on the right-hand side
with $S_{fp}(\sigma) \not= \emptyset$ represents a set of $\prod_{j\in S_{fp}}z_j$
distinct
spanning digraphs which have in common the specified cycles $c_1,\dots , c_p$,
and differ in the oriented edges outgoing from the vertices
$j\in S_{fp}(\sigma)$. These oriented edges may form cycles on
their own which do not enter into the list $c_1,\dots , c_p$; (iii) Since
the sets $\cup_{i=1}^p\O_i$ and $S_{fp}(c_1,\dots , c_p)$ are disjoint,
the proper cycles formed by the oriented edges incident to the fixed points of
a given permutation $\sigma = c_1\circ c_2 \circ \cdots \circ c_p$ should
enter into the enlarged list of cycles $c_1, c_2, \dots, c_p,\dots, c_{p'}$,
$p'>p$, corresponding to the cycle decomposition of another permutation $\sigma'$.

\begin{figure}[h!]
\includegraphics[width=60mm, height=70mm]{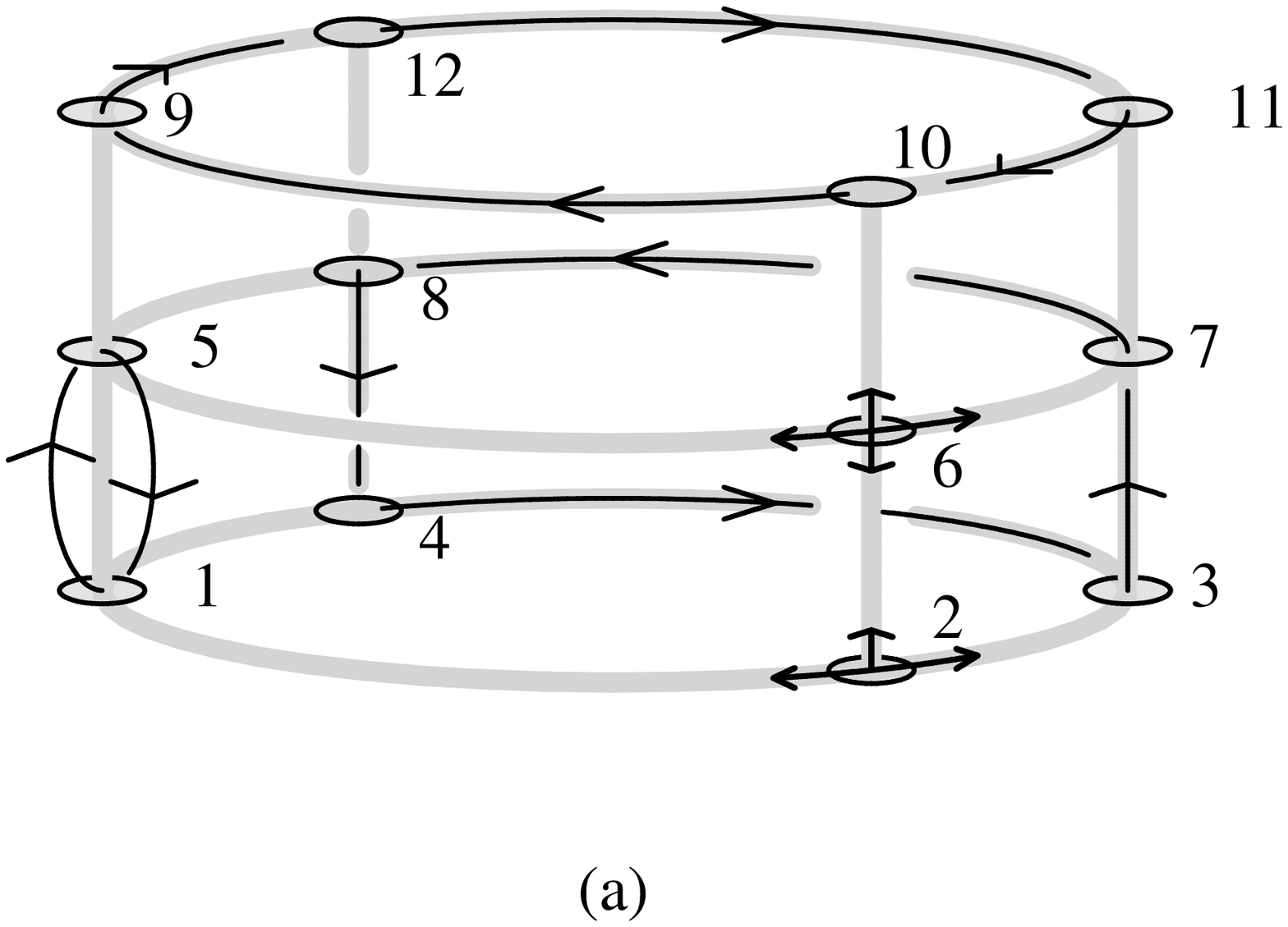}\vspace*{-1cm}
\includegraphics[width=60mm, height=70mm]{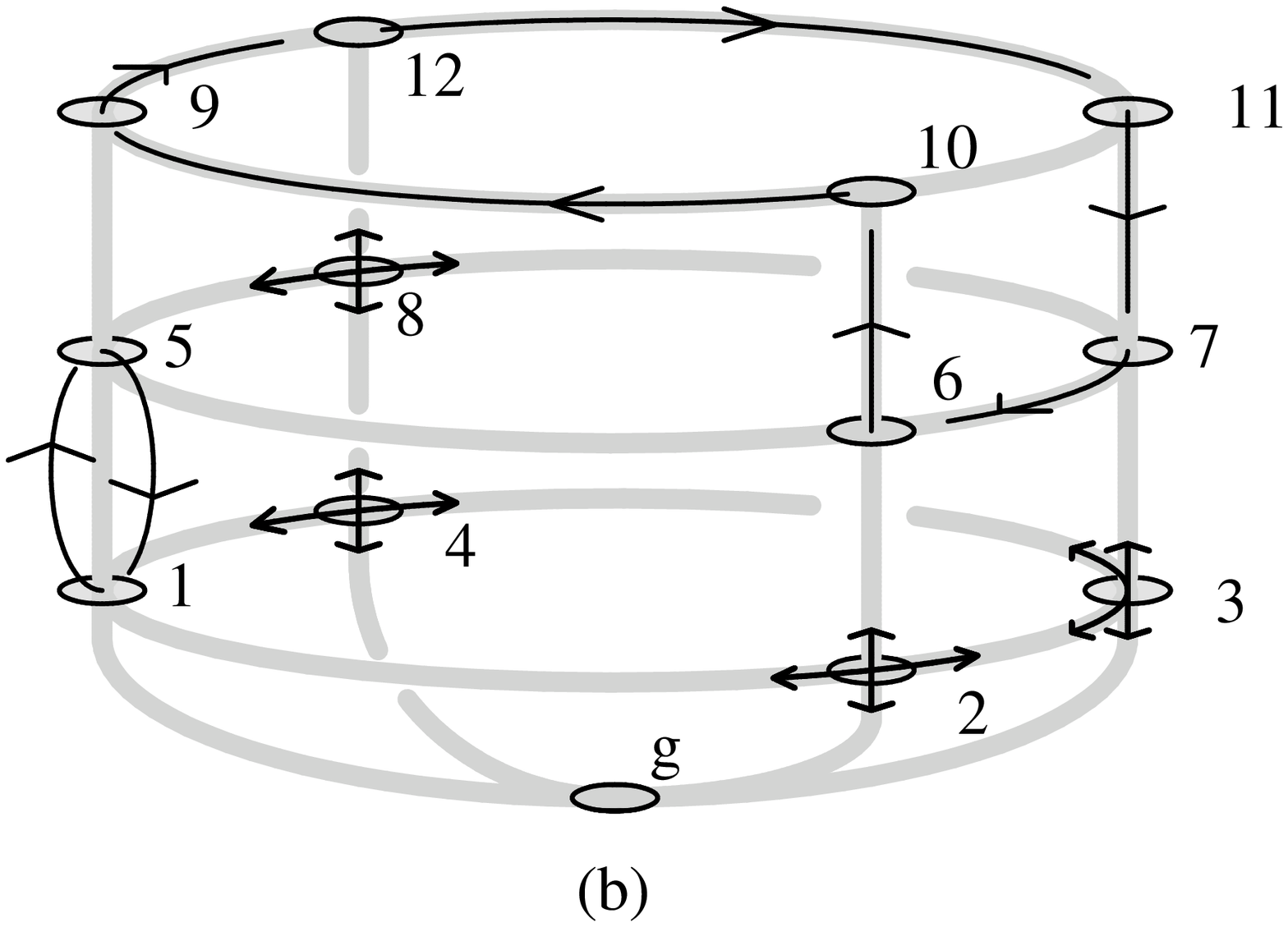}\vspace*{1cm}
\caption{\label{fig1} Possible spanning digraphs on a cylinder, generated by a single term in the
determinant expansion of the corresponding Laplace matrix (see text). Part
(a) corresponds to closed boundary conditions, and part (b) to open at the bottom
and closed at the top ones.}
\end{figure}

For example, consider the determinant of the Laplacian matrix of a cylinder of height 3
and perimeter 4. In the case of closed boundary conditions, the
corresponding Leibniz expansion
will contain the term
\begin{equation}
\label{LeibnExam1}
(-1)^3 (\Delta_{1,5}\Delta_{5,1})\Delta_{2,2}(\Delta_{3,7}\Delta_{7,8}\Delta_{8,4}\Delta_{4,3}) %
       \Delta_{6,6}(\Delta_{9,12}\Delta_{12,11}\Delta_{11,10}\Delta_{10,9}),
\end{equation}
which represents, up to the sign, $z_{2}z_{6}= 12$ spanning digraphs on $\G$ with 3 specified
cycles and all possible oriented bonds outgoing from the vertices 2 and 6, as illustrated in
Fig. \ref{fig1} (a). One of these 12 spanning digraphs will contain the additional cycle
$2 \rightarrow 6 \rightarrow 2$.
In the case of open boundary at the bottom and closed at the top, the term
\begin{equation}
\label{LeibnExam2}
(-1)^2 (\Delta_{1,5}\Delta_{5,1})\Delta_{2,2}\Delta_{3,3}\Delta_{4,4}(\Delta_{6,10}\Delta_{10,9}  %
       \Delta_{9,12}\Delta_{12,11}\Delta_{11,7}\Delta_{7,6})\Delta_{8,8}
\end{equation}
represents $z_{2}z_{3}z_4 z_8 = 256$ spanning digraphs on $\G'$ with 2 specified
cycles and all possible oriented bonds outgoing from the vertices 2, 3, 4 and 8, as illustrated in
Fig. \ref{fig1} (b). The latter bonds will generate three digraphs with one additional cycle of length 2:
$2 \rightarrow 3 \rightarrow 2$, or $3 \rightarrow 4 \rightarrow 3$, or $4 \rightarrow 8 \rightarrow 4$,
and one digraph with two additional cycles, $2 \rightarrow 3 \rightarrow 2$ and $4 \rightarrow 8
\rightarrow 4$.

As noticed first in \cite{Pr85}, the expansion (\ref{altern}) parallels in form the
\textit{inclusion-exclusion principle} in combinatorial mathematics. Indeed, let
$c_1, c_2, \dots, c_m$ be the list of all possible proper cycles on $\G$,
labeled in an arbitrary order. Define $A_i$, $i=1,2,\dots ,m$ as
the set of all spanning digraphs on $\G$ containing the particular cycle $c_i$.
Then, expansion (\ref{altern}) can be written in the form of the inclusion-exclusion
principle:
\begin{equation}
\left| \cup_{i=1}^m A_i \right|= \sum_{i=1}^m
|A_i| - \sum_{1\leq i<j \leq m} |A_i \cap A_j| + \sum_{1\leq i<j<k
\leq m} |A_i\cap  A_j \cap A_k|- \cdots +(-1)^{m+1}|A_1 \cap
\cdots \cap A_m|, \label{inex}
\end{equation}
which holds for any finite sets $A_1, A_2, \dots , A_m$,
where $|A|$ is the cardinality of the set $A$.

Now we are in the position to define a matrix $D^{(0,0)}$, associated with the graph
$\G$, such that $\det D^{(0,0)}$ be the generating function of all spanning digraphs
on $\G$ which have no contractible cycles. The elements $D_{ij}^{(0,0)},
i,j=1,\dots, n=NM$ of $D^{(0,0)}$ are explicitly given as:
\begin{equation}\label{laplacian omega}
D_{ij}^{(0,0)} = \left\{ \begin{array}{rll}
z_i,& \mathrm{if}& \, i=j, \\
-1,&  \mathrm{if}& \, i,j \,\, \mathrm{are} \, \, \mathrm{vertical}\, \, \mathrm{neighbors},\\
-a,&  \mathrm{if}& \, i \,\, \mathrm{is} \, \, \mathrm{left}\, \, \mathrm{neighbor}
\,\, \mathrm{of} \, \, j,\\
- a^{-1},&  \mathrm{if}& \, i \,\, \mathrm{is} \, \, \mathrm{right}\, \, \mathrm{neighbor}
\,\, \mathrm{of} \, \, j,\\
0,& &\mathrm{otherwise}.
\end{array}      \right.
\end{equation}

Here $a= \omega^{1/M} e^{-i\pi/M}$, the terms "left" and "right"
fix the opposite directions of the horizontal edges. Note that all
closed paths which do not wrap the cylinder contain an equal
number of horizontal edges with either orientations, hence, their
weight in $\det D^{(0,0)}$ remains the same as in $\det\Delta$.
Therefore, all the configurations which contain such closed paths
(contractible cycles) cancel out in the expansion of $\det
D^{(0,0)}$. On the other hand, cycles generated by off-diagonal
elements that wrap the cylinder change their sign, because they
contain horizontal edges oriented in one direction exceeding by
$M$ the number of edges in the opposite direction. This amounts to
the total factor of $a^M=-\omega$, or $a^{-M}=-\omega^{-1}$
depending on the orientation. Therefore, each non-contractible
cycle with a given orientation is counted twice, however, with
different weight - once it enters into the determinant expansion
with unit weight, being generated by diagonal elements of the
matrix $D^{(0,0)}$, and second time it enters with a factor
$\omega$ or $\omega^{-1}$ (depending on the orientation) as
generated by off-diagonal elements of that matrix. Thus, the total
number of non-contractible cycles, irrespective of their origin
and orientation, is given by the coefficient in front of the
corresponding power of $\omega + \omega^{-1} +2 \equiv \xi$ in the
series expansion of the partition function. The power of $\xi$ is
the "good quantum number" which is a well defined and conserved
quantity under the action of the transfer matrix along the
cylinder. However, as we shall see below, more convenient
expansions of the partition function, which can be directly
compared with characters of the Virasoro modules, are given by the
power series in $\omega$ itself, or in terms of combinations like
$(\omega + \omega^{-1})^s$ and $\sum_{k=0}^s \omega^{s-2k}$. In
general, besides the non-contractible cycles, the spanning digraph
contains tree subgraphs connected to them. All branches of the
trees can be generated only by the diagonal elements of
$D^{(0,0)}$ and, hence, carry unit weight.

\subsection{Cylinder with one closed and one open boundary}

\begin{figure}[h!]
\includegraphics[width=80mm]{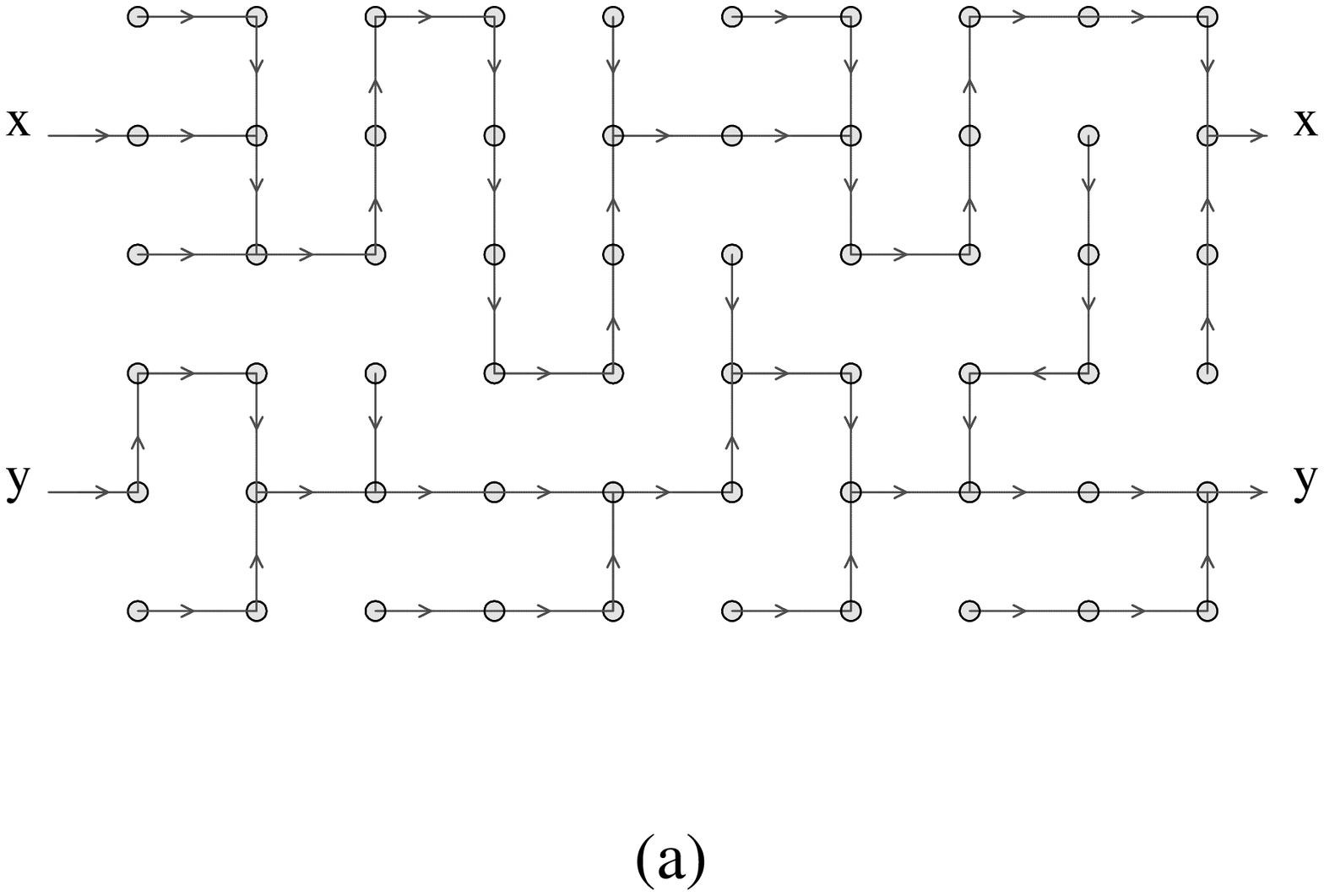}
\includegraphics[width=80mm]{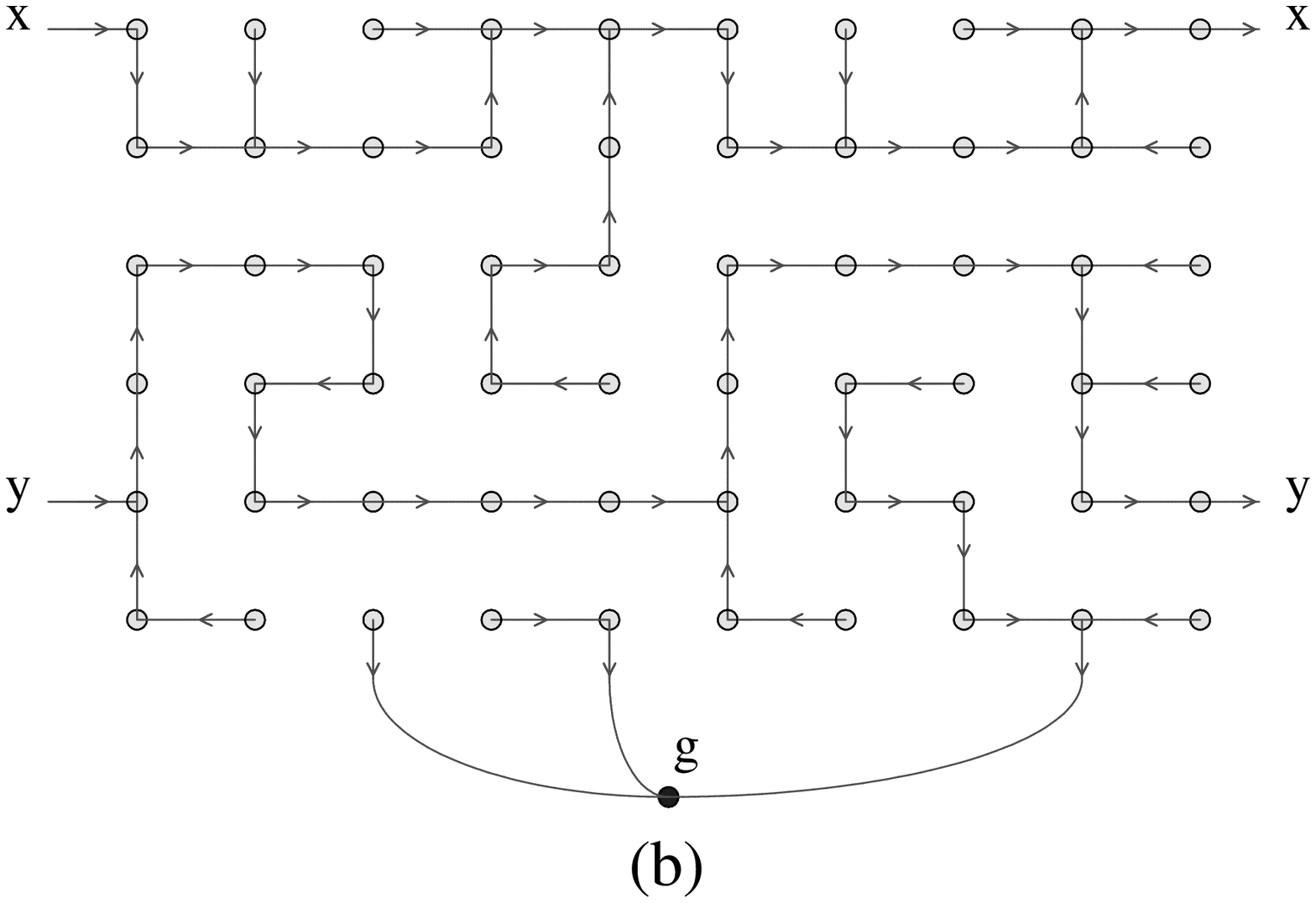}
\caption{\label{loopTree} Spanning digraphs on a cylinder of height $N=6$ and
perimeter $M=10$ under:
(a) Closed  boundary conditions. The two closed paths $x-x$ and $y-y$ represent
non-contractible cycles. (b) Open boundary conditions at the bottom and closed ones
at the top. The three bonds incident with the root give rise to boundary trees.}
\end{figure}

The Laplacian matrix $\Delta^{\star}$ for the graph $\G'$ corresponding to $(\mu,\nu)=(0,1)$
boundary conditions is a $(n+1)\times(n+1)$
matrix of the same form (\ref{laplacian}) as far as the notions of degree of
a vertex and adjacency are understood in the context of $\G'$. However,
to make the similarities and dissimilarities with the former case
apparent, we retain the notation $z_i$ for the degree of vertex $i$, $i=1,\dots ,n$,
with respect to $\G$, and explicitly introduce the label $g$ for the root with degree
$M$ in $\G'$, see Fig. (\ref{loopTree}) b. Thus, for the matrix elements
of $\Delta^{\star}$ we have
\begin{equation}
\label{Lapstar}
\Delta^{\star}_{ij} = \left\{ \begin{array}{rll}
z_i,& \mathrm{if}& \, i=j\in V\setminus B, \\
z_i+1,& \mathrm{if}& \, \, i=j\in B, \\
M,& \mathrm{if}& \, \, i=j=g, \\
-1,&  \mathrm{if}& \, \, i,j\in V \,\, \mathrm{are} \, \, \mathrm{adjacent}\, \,
\mathrm{in} \, \, \G,\\
-1,&  \mathrm{if}& \, \, i\in B, j=g \,\, \mathrm{or} \, \, i=g, j\in B,\\
0,&& \mathrm{otherwise}.
\end{array}      \right.
\end{equation}

Here $B$ is the set of bottom boundary vertices adjacent to the root $g$ in $\G'$.
Now we make use of the fact that by Kirchhoff's theorem the number $N_{st}$ of
spanning trees on $\G'$ is equal to any cofactor
of $\Delta^{\star}$ and chose for convenience the cofactor $C_{gg}$ of the element
$\Delta^{\star}_{gg}$. Then $N_{st}= C_{gg}=\det \Delta'$, where $\Delta'$ is the
$n\times n$ matrix with elements ($i,j=1,\dots ,n$):
\begin{equation}\label{Lapprime}
\Delta'_{ij} = \left\{ \begin{array}{rll}
z_i,& \mathrm{if}& \, i=j\in V\setminus B, \\
z_i+1,& \mathrm{if}& \, \, i=j\in B, \\
-1,&  \mathrm{if}& \, \, i,j\in V \,\, \mathrm{are} \, \, \mathrm{adjacent},\\
0,&& \mathrm{otherwise}.
\end{array}      \right.
\end{equation}

By comparing the above expression with (\ref{laplacian}) one sees that the
only difference is
in the diagonal elements: the order of the vertices belonging to the open
boundary has been increased by one. Therefore, the same arguments as in the
previous section lead us to the matrix $D^{(0,1)}$ with elements
\begin{equation}\label{Dprime}
D^{(0,1)}_{ij} =\left\{
\begin{minipage}{4.5cm}
$\text{\ \ }  z_i, \qquad  \text{if $i=j\in V\setminus B$},$\\
$            z_i+1,  \qquad   \text{ if $i=j\in B$},$ \\
$            -1,  \qquad   \text{if $i,j$ are vertical neighbors},$ \\
$            -a,  \qquad   \text{if $i$ is left neighbor of $j$},$ \\
$            -a^{-1},  \qquad   \text{if $i$ is right neighbor of $j$},$ \\
$\text{\ \ }  0,  \qquad   \text{otherwise}.$
\end{minipage}     \right.
\end{equation}
where, as before, $a= \omega^{1/M} e^{-i\pi/M}$. By construction, in
the expansion of $\det D^{(0,1)}$ all the arrow configurations with
contractible cycles cancel out. Thus $\det D^{(0,1)}$ is the generating
function of all the spanning digraphs which are either spanning
forests of trees rooted at the open bottom boundary, or contain
non-contractible cycles wrapping the cylinder and tree subgraphs
rooted at these cycles or at the open boundary. As in the case of
closed boundaries, each non-contractible cycle with a given
orientation is counted twice with different weights - once with unit
weight, being generated by diagonal elements of the matrix $D^{(0,1)}$,
and second time with a factor $\omega$ or $\omega^{-1}$ (depending on
the orientation) as generated by off-diagonal elements of that
matrix.

The case of open-open boundary conditions, $(\mu,\nu)=(1,1)$, can be
considered analogously to the previous two cases and we do not describe
it separately.

The partition function calculated as the determinant of the Laplacian,
$Z^{(\mu,\nu)}_{NM}=\det D^{(\mu,\nu)}$,
can be split into a product of two parts,
\begin{equation}\label{universal}
Z^{(\mu,\nu)}_{NM}=e^{F^{(\mu,\nu)}_{NM}}\bar{Z}^{(\mu,\nu)}_{N}(q,\omega)
\end{equation}
where $e^{F^{(\mu,\nu)}_{NM}}$ is the {\it nonuniversal} part of
the partition function, including bulk and boundary free energy
$F^{(\mu,\nu)}_{NM}$, and $\bar{Z}^{(\mu,\nu)}_{N}(q,\omega)$ is
the {\it universal} part which is a polynomial in the aspect ratio
parameter $q=e^{-\frac{\pi M}{N}}$ and a Laurent polynomial in
$\omega$. The universal part of the partition function can be
decomposed into different combinations of $\omega$, for example,
\begin{eqnarray}
\label{N polynom1}
\bar{Z}^{(\mu,\nu)}_{N}(q,\omega) &=&\sum_{s=0}^N C^{(\mu,\nu)}_{1,s}[N](q)
(\omega+\omega^{-1})^s, \\
\label{N polynom2}
\bar{Z}^{(\mu,\nu)}_{N}(q,\omega) &=&\sum_{s=-N}^N C^{(\mu,\nu)}_{2,s}[N](q) \omega^s, \\
\label{N polynom3}
\bar{Z}^{(\mu,\nu)}_{N}(q,\omega) &=&   \sum_{s=0}^N C^{(\mu,\nu)}_{3,s}[N](q)
\sum_{k=0}^{s}\omega^{s-2k},
\end{eqnarray}
valid for all boundary conditions under consideration, $\mu, \nu
=0,1$. Note that the second and third polynomials can be obtained
from the first one by using the relations:
\begin{eqnarray}
\label{C polynom1}
C^{(\mu,\nu)}_{2,s}[N]&=&C^{(\mu,\nu)}_{2,-s}[N]=\sum_{p=0}^{\left[\frac{N-s}{2}\right]}
{{s+2p} \choose {p}} C^{(\mu,\nu)}_{1,s+2p}[N], \quad  s=0,\ldots,N ,\\
\label{C polynom2}
C^{(\mu,\nu)}_{3,s}[N]&=&C^{(\mu,\nu)}_{2,s}[N]-C^{(\mu,\nu)}_{2, s+2}[N], \quad s=0,
\ldots,N-2 ,\\
\label{C polynom3}
C^{(\mu,\nu)}_{3,s}[N]&=&C^{(\mu,\nu)}_{2,s}[N], \quad  s=N-1,  N.
\end{eqnarray}

Each of the polynomials $C^{(\mu,\nu)}_{i,s}[N](q)$, $i=1,2,3$, has a
well defined statistical meaning. For example, $C^{(\mu,\nu)}_{1,s}[N](q)$
is the universal factor in the partition function which is proportional to the
number of configurations with $s$ non-contractible cycles of either orientation,
generated by off-diagonal elements of the matrix $D^{(\mu,\nu)}$; these configurations
may have also any allowed (by the size of the cylinder) number of non-contractible
cycles of both orientations, generated by diagonal elements of that matrix. On the
other hand, the universal factor $C^{(\mu,\nu)}_{2,s}[N](q)$ is proportional to the
number of configurations with fixed to $s$ \emph{difference in the numbers} of
non-contractible cycles with positive and negative
orientation, generated by off-diagonal elements of the matrix $D^{(\mu,\nu)}$; the
total number of such cycles, as well as the number of non-contractible
cycles of either orientation, generated by diagonal elements of the same matrix, may
take any allowed values.

In the next section, we shall evaluate the asymptotic form of the
coefficients $C^{(\mu,\nu)}_{i,s}[N]$, $i=2,3$, for $M\rightarrow
\infty$, $N\rightarrow \infty$ and disclose their relation to the
finitized characters of the logarithmic Virasoro modules.

\section{Calculation of the partition function}

The matrices of edge weights $D^{(\mu,\nu)}$ can be written in the
form of a sum of direct products of
simple $N\times N$ and $M\times M$ matrices:
\begin{equation}
D^{(\mu,\nu)} = (2E_N - Q^{(\mu,\nu)}_{N})\otimes E_M +  E_N\otimes
(2E_M +a G_M + a^{-1} G^T_M),
 \label{matrix}
\end{equation}
where $E_N$ ($E_M$) is the unit $N\times N$ ($M\times M$) matrix,
$Q^{(\mu,\nu)}_{N}= \{q^{(\mu,\nu)}_{i,j}\}$ is a tridiagonal matrix
with elements dependant on the boundary conditions ($i,j=1,\dots ,N$):
\begin{eqnarray}
q^{(0,0)}_{i,j}&=&\delta_{i,1}\delta_{j,1}+\delta_{i-1,j}+
\delta_{i,j-1}+
\delta_{i,N}\delta_{j,N}, \\
q^{(0,1)}_{i,j}&=&q^{(1,0)}_{i,j}=\delta_{i,1}\delta_{j,1}+\delta_{i-1,j}+
\delta_{i,j-1}, \\
q^{(1,1)}_{i,j}&=&\delta_{i-1,j}+ \delta_{i,j-1}.   \label{elements}
\end{eqnarray}

Note that $2E_N - Q^{(\mu,\nu)}_{N}$ differs only by sign from the
one-dimensional discrete Laplacian on a chain of $N$ sites with the
analogues of Neumann-Neumann ($\mu=\nu =0$), Neumann-Dirichlet
($\mu=0,\nu =1$ or $\mu=1,\nu =0$) and Dirichlet-Dirichlet
($\mu=1,\nu =1$) boundary conditions. Hence, the eigenvalues of
$2E_N - Q^{(\mu,\nu)}_{N}$ are $\lambda_N^{(0,0)}(p) = 2 - 2\cos
\frac{\pi p}{N}$, $\lambda_N^{(0,1)}(p)=\lambda_N^{(1,0)}(p) = 2 -
2\cos \frac{\pi (2p+1)}{2N+1}$, and $\lambda_N^{(1,1)}(p) = 2 -
2\cos \frac{\pi (p+1)}{N+1}$, where $p=0,\dots, N-1$. The $M\times M$
matrix $G_M= \{g_{m,n}\}$ in (\ref{matrix}) has the elements
$g_{m,n}=\delta_{m-1,n}+ \delta_{m,M}\delta_{n,1}$, with
$m,n=1,\dots ,M$, and $G^T_M$ is the transposed of $G_M$. Their
eigenvalues are $\mu_M(k) = \mathrm{exp}(\mathrm{i} 2\pi k/M)$ and
$\bar{\mu}_M(k) = \mathrm{exp}(-\mathrm{i} 2\pi k/M)$, respectively,
where $k=0,\dots ,M-1$. Thus, for the corresponding partition
functions $Z_{NM}^{(\mu,\nu)} = \det D^{(\mu,\nu)}_{NM}$ we obtain:
\begin{equation}
Z_{NM}^{(\mu,\nu)} = \prod_{p=0}^{N-1}\prod_{k=0}^{M-1}
\left[\lambda^{(\mu,\nu)}_N(p) +2 -
\omega^{1/M}\mathrm{e}^{\mathrm{i}\pi (2k+1)/M} -
\omega^{-1/M}\mathrm{e}^{-\mathrm{i}\pi (2k+1)/M}\right] .
 \label{PFtau}
\end{equation}

By analytic continuation of the identity
\begin{equation}
\prod_{k=0}^{M-1}\left[Q^2+Q^{-2}-
2\cos\left(\frac{2\pi k}{M}+\alpha \right)\right]=
Q^{2M}+ Q^{-2M}-2 \cos(M\alpha),
\label{iden}
\end{equation}
with
\begin{eqnarray}
&& Q \equiv Q_N^{(\mu,\nu)}(p) = \sqrt{1+\sin^2 \phi_N^{(\mu,\nu)}(p)}+ \sin
\phi_N^{(\mu,\nu)}(p), \\ && \nonumber
\phi_N^{(0,0)}(p)=\frac{\pi p}{2N}, \quad
\phi_N^{(0,1)}(p)=\phi_N^{(1,0)}(p)= \frac{\pi (2p+1)}{2(2N+1)},\quad
\phi_N^{(1,1)}(p)=\frac{\pi (p+1)}{2(N+1)}, \label{param}
\end{eqnarray}
from real $\alpha$ to complex $\alpha =(\pi - \mathrm{i}\ln\omega)/M$,
we obtain
\begin{equation}\label{ZNM}
Z_{NM}^{(\mu,\nu)} = \prod_{p=0}^{N-1} [Q_N^{(\mu,\nu)}(p)]^{2M}\left\{
1+(\omega +\omega^{-1})[Q_N^{(\mu,\nu)}(p)]^{-2M}+
[Q_N^{(\mu,\nu)}(p)]^{-4M}\right\}. \label{Ztau}
\end{equation}
The above exact partition function is a polynomial in
$x=\omega +\omega^{-1}$ of degree $N$, see (\ref{N polynom1}).

When $M,N \rightarrow \infty$, so that $M/N=O(1)$, a standard finite-size
analysis of
the contribution to the free energy from the $\omega$-independent
factor in the partition function yields the asymptotic expansion
\begin{eqnarray}
&&\ln \prod_{p=0}^{N-1} [Q_N^{(\mu,\nu)}(p)]^{2M} \simeq
\nonumber \\ &&\frac{4G}{\pi}MN +
M\left[\frac{2G}{\pi}(\mu+\nu) - \ln(1+\sqrt{2})\right]
-\frac{\pi M}{N}\left[\frac{1}{12}-\frac{(\mu-\nu)^2}{8}\right],
\label{Stan1}
\end{eqnarray}
where $G$ is Catalan's constant.

\begin{figure}[h!]
\includegraphics[width=80mm]{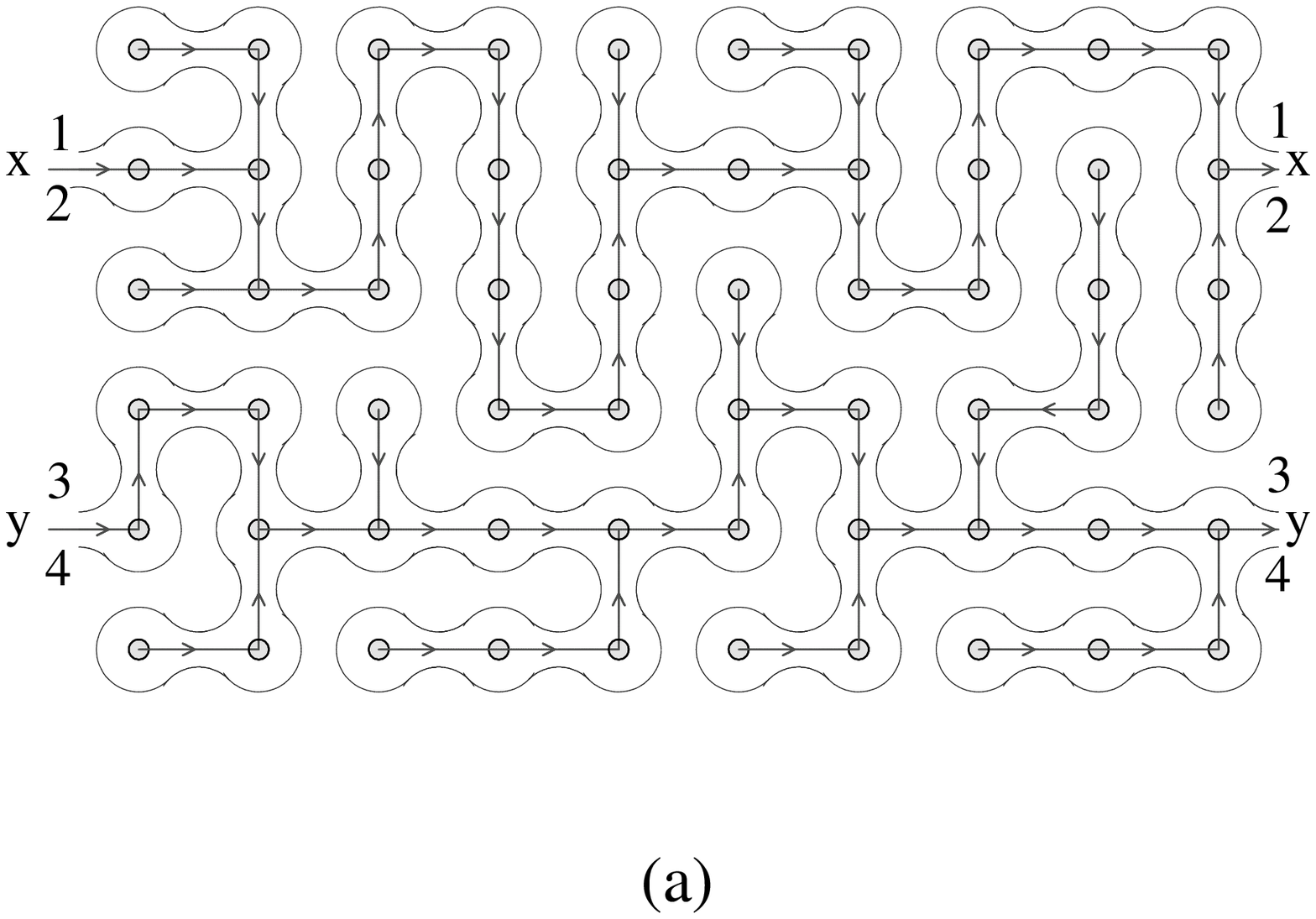}
\includegraphics[width=80mm]{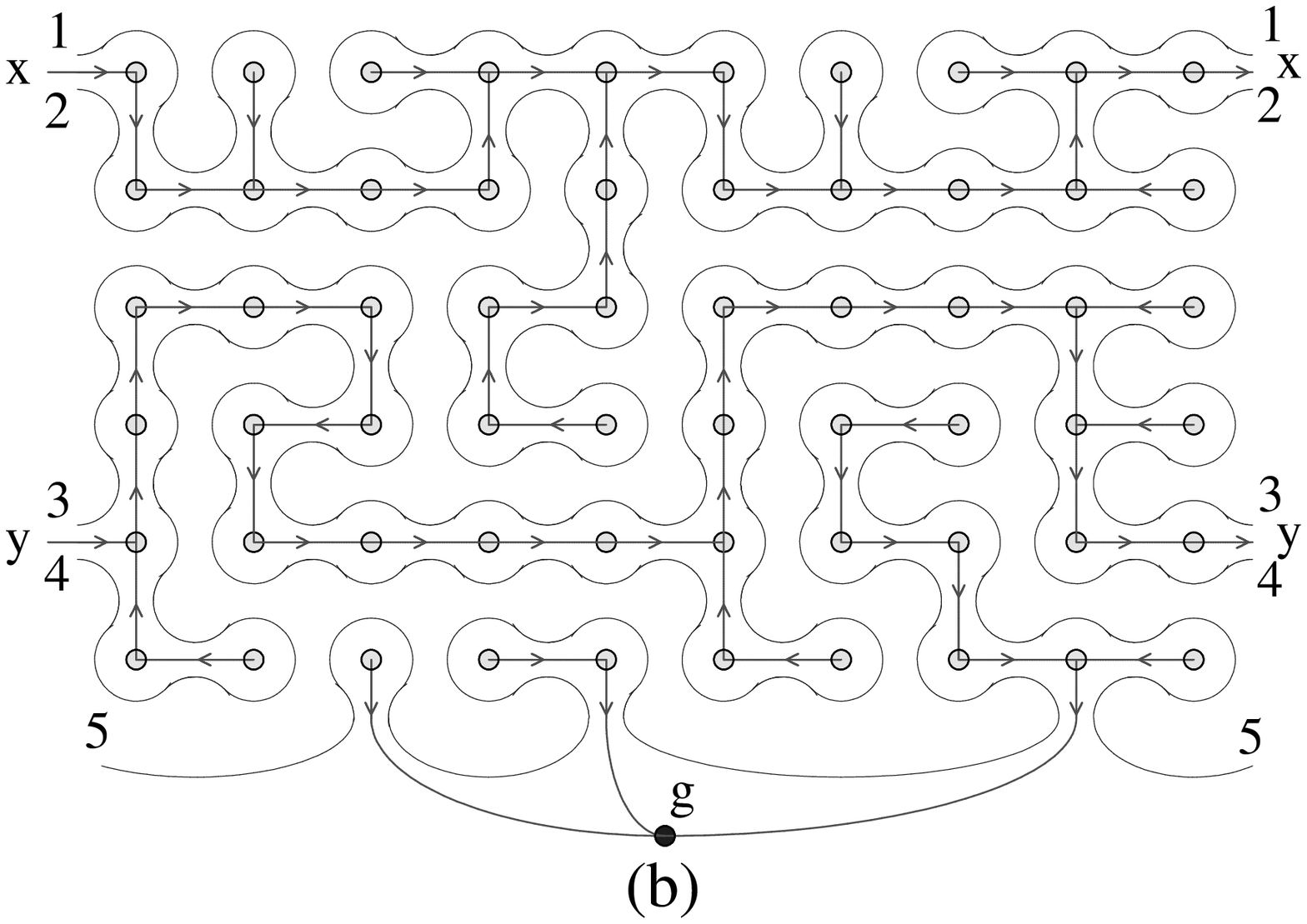}
\caption{\label{densed} Correspondence between spanning webs and
dense polymers. (a) The cycle $x-x$ is surrounded by two polymers, $1-1$
and $2-2$, and the cycle $y-y$ by another two polymers, $3-3$ and $4-4$,
which are considered as defect lines in the classification
of \cite{PRZ, PearceRas}. (b) Besides the four defect polymer lines, there
is a fifth line $5-5$ separating the boundary trees from the rest of the
lattice.}
\end{figure}

The leading-order asymptotic form of the $\omega$-dependent factor
in the partition function follows from the approximation
$Q_N^{(\mu,\nu)}(p)\simeq 1+\phi_N^{(\mu,\nu)}(p)$.
Thus we obtain
\begin{equation}
\prod_{p=0}^{N-1}\left\{ 1+x[Q_N^{(\mu,\nu)}(p)]^{-2M}+
[Q_N^{(\mu,\nu)}(p)]^{-4M}\right\} \simeq \prod_{p=0}^{N-1}\left\{
1+x\mathrm{e}^{-\frac{\pi M}{2N}(2p+\mu+\nu)} +\mathrm{e}^{-\frac{\pi
M}{N}(2p+\mu+\nu)}\right\}, \label{omdep}
\end{equation}
where $x=\omega + \omega^{-1}$.

For $N$ and $M$ large and $N/M$ fixed, we rewrite (\ref{ZNM}) in
the form (\ref{universal}) with the free energy
\begin{equation}
F^{(\mu,\nu)}_{MN}=\frac{4G}{\pi}MN
 +M\left[\frac{2G}{\pi}(\mu+\nu) - \ln(1+\sqrt{2})\right],
\end{equation}
and \emph{universal part} of the partition function
\begin{equation}\label{conf-ZNM}
\bar{Z}_{N}^{(\mu,\nu)}(q,\omega)=q^{\frac{1}{12}-\frac{1}{8}
(\mu-\nu)^2}\prod_{j=0}^{N-1}\left[1+(\omega+\omega^{-1})
q^{j+\frac{1}{2}(\mu+\nu)}+q^{2j+\mu+\nu}\right].
\end{equation}
In the next section, we show that (\ref{conf-ZNM}) converges as
$N\to\infty$ to the characters of symplectic fermions \cite{Kausch}
or, equivalently, to the characters of the doublet algebra
$\mathcal{A}(2)$ \cite{FFT} in the $p=2$ logarithmic model. As is
well known, the
characters contain complete information about conformal dimensions
and their multiplicities in the model.

Before considering the symplectic fermions, we calculate the coefficients
$C^{(\mu,\nu)}_{2,s}[N](q)$
(\ref{C polynom2}) and $C^{(\mu,\nu)}_{3,s}[N](q)$ (\ref{C polynom3})
explicitly.
The universal part of the partition function (\ref{conf-ZNM}) can be
rewritten by using the Newton $q$-binomial formula
\begin{equation}
\prod_{p=0}^{N-1}(1+y q^p)=\sum_{s=0}^N
q^{\frac{s(s-1)}{2}} {N \choose s}_q y^s,
\label{Newton}
\end{equation}
with $q$-binomial coefficients
\begin{eqnarray}
{N \choose s}_q=\frac{(1-q^N)\cdots
(1-q^{N-s+1})}{(1-q)\cdots (1-q^s)},\quad \mbox{when}
\quad 0\leq s\leq N, \nonumber \\
{N \choose s}_q=0,\quad \mbox{when}\quad s<0\quad \mbox{or}\quad s>N,
\label{Binomial}
\end{eqnarray}
in the form (\ref{N polynom2}) with
\begin{equation}
C^{(\mu,\nu)}_{2,s}[N](q)= q^{\frac{1}{12}-\frac{(\mu -\nu)^2}{8}+\frac{s(s-1+\mu+\nu)}{2}}
\sum_{k=0}^{N-s}q^{k^2+k(s-1+\mu+\nu)} {N \choose k}_q{N \choose s+k}_q, \quad 0 \leq s\leq N,
\label{C2s}
\end{equation}
and $C^{(\mu,\nu)}_{2,s}[N](q)= C^{(\mu,\nu)}_{2,-s}[N](q)$.

For closed-closed and open-closed boundary conditions the summation can be performed
explicitly and the above expression simplifies to
\begin{equation}
C^{(0,0)}_{2,s}[N](q)=q^{\frac{1}{12}+\frac{s(s-1)}{2}}\left[
{2N \choose N+s-1}_q + q^{s} {2N \choose N+s}_q \right]
\label{C2s-00s}
\end{equation}
and
\begin{equation}
C^{(0,1)}_{2,s}[N](q)=q^{-\frac{1}{24}+\frac{s^2}{2}}
{2N \choose N+s}_q ,
\label{C2s-01s}
\end{equation}
respectively.

Now we give explicit expressions for
$C^{(\mu,\nu)}_{3,s}[N](q)$, which are shown in the next section
to converge to the Virasoro characters as $N\to\infty$. Using
(\ref{C polynom2}), we obtain for $0\leq s \leq N-2$:
\begin{eqnarray}
C^{(\mu,\nu)}_{3,s}[N](q)&=&C^{(\mu,\nu)}_{2,s}[N](q)-C^{(\mu,\nu)}_{2,s+2}[N](q)
\nonumber \\ &=&q^{\frac{1}{12}-\frac{(\mu
-\nu)^2}{8}+\frac{s(s-1+\mu+\nu)}{2}}\left\{
\sum_{k=0}^{N-s}q^{k^2+k(s-1+\mu+\nu)} {N \choose k}_q{N \choose
s+k}_q - \right.\nonumber \\
&-&\left. q^{s+1} \sum_{k=1}^{N-s-1}q^{k^2+k(s-1+\mu+\nu)} {N
\choose k-1}_q{N \choose s+k+1}_q \right\} .
\end{eqnarray}

We note that
\begin{eqnarray}\label{C300}
&& C^{(0,0)}_{3,s}[N](q)= q^{\frac{1}{12}}\left[q^{\frac{s(s-1)}{2}}
{2N-1 \choose N+s+1}_q + q^{\frac{s(s+1)}{2}}{2N-1 \choose N+s}_q
\right. \nonumber \\ && \left. - q^{\frac{(s+1)(s+2)}{2}}{2N-1\choose N+s+1}_q
- q^{\frac{(s+2)(s+3)}{2}}{2N-1 \choose N+s+2}_q \right]=
\chi_{2s+1}^{(2N-1)}(q) + \chi_{2s+3}^{(2N-1)}(q),
\end{eqnarray}
where $\chi_{s}^{(N)}(q)$ are the finitized characters of the logarithmic Virasoro
modules, given by Eq. (5.5) in Ref. \cite{PearceRas} for $s$ odd.
It should be mentioned that (\ref{C300}) resembles Eq. (7.25) in Ref. \cite{PearceRas}
for the finitized character of the logarithmic Virasoro module.

Similarly,
\begin{equation}
C^{(0,1)}_{3,s}[N](q)= q^{-\frac{1}{24}+\frac{s^2}{2}}\left[
{2N \choose N+s}_q - q^{2s+2}{2N \choose N+s+2}_q\right]=
\chi_{2s+2}^{(2N+1)}(q)
\label{C3-01}
\end{equation}
coincides with the finitized Virasoro characters $\chi_{2s+2}^{(2N+1)}(q)$ given by
Eq. (5.5) in Ref. \cite{PearceRas} for $s$ even.

On the other hand, under the substitution $(N,s)\rightarrow(2N+1,2s+1)$ in
Eq. (5.5) in Ref. \cite{PearceRas} for $s$ odd, we obtain the relationship
\begin{equation}
\chi_{2s+1}^{(2N+1)}(q)= C^{(1,1)}_{3,s-1}[N](q)+C^{(1,1)}_{3,s}[N](q) .
\end{equation}
This relation corresponds to the fact that $\chi_{2s+1}^{(2N+1)}(q)$ is the finitized character
of the rank-1
indecomposable Virasoro module with two irreducible subqotients whose finitized characters
are given by $C^{(1,1)}_{3,s-1}[N](q)$ and $C^{(1,1)}_{3,s}[N](q)$.

Next, Eq. (\ref{conf-ZNM}) implies
\begin{equation}\label{ZN0011}
\bar{Z}_{N+1}^{(0,0)}(q,\omega)=(\omega+2+\omega^{-1})\bar{Z}_{N}^{(1,1)}(q,\omega).
\end{equation}
Hence, the coefficients $C^{(0,0)}_{3,s}[N]$ and $C^{(1,1)}_{3,s}[N-1]$ are related by
the equality
\begin{equation}\label{rel0011}
 C^{(0,0)}_{3,s}[N](q)=C^{(1,1)}_{3,s-1}[N-1](q)+2C^{(1,1)}_{3,s}[N-1](q)
 +C^{(1,1)}_{3,s+1}[N-1](q),
\end{equation}
which repeats the relations between characters of the logarithmic and
irreducible Virasoro modules.

Finally, from Eqs. (\ref{C2s-00s}) and (\ref{C2s-01s}) it follows that the conformal
wights for the open-open and open-closed boundary conditions are
\begin{equation}\label{W00}
\Delta_s^{(0,0)}= \frac{s(s-1)}{2} , \quad s=0,1,2,\dots
\end{equation}
and
\begin{equation}\label{W01}
\Delta_s^{(0,1)}= \frac{4s^2-1}{8} , \quad s=0,1,2,\dots
\end{equation}
These two sequences can be arranged into the first column of the extended Kac table so
that its odd entries correspond to (\ref{W00}) and the even ones to (\ref{W01})

It is instructive to compare the obtained results with those of
Pearce and Rasmussen \cite{PearceRas}. The partition function of
dense polymers evaluated in \cite{PearceRas} is a function of the
number of defect lines $l=0,1,2,\dots$ in the polymer system, so
that the extended Kac label $s=l+1$ runs over all entries in the
first column of the table of conformal weights. It is easy to
notice that even cells differ from their odd counterparts by a
fixed value $-1/8$. This value can be associated with the
conformal weight $h_{min}$ of the operator with the smallest
scaling dimension present in the spectrum of the Hamiltonian,
which depends on the boundary conditions (see for instance
\cite{IPRH} where it is shown that $h_{\rm{min}}=0$ for open-open
boundary conditions and $h_{\rm{min}}=-1/8$ for open-closed ones).
However, for the dense polymers model the boundary conditions at
both sides of the infinite strip are equal for even and odd values
of $s$. A comparison with the model of spanning webs resolves this
illusive contradiction. The correspondence between our spanning
webs model and the model of critical dense polymers is shown in
Fig. \ref{densed}. The left-hand figure shows a spanning web for
closed boundary conditions. The polymer lines envelop branches and
cycles of the spanning web so that each cycle is surrounded by two
polymer lines. Then, for closed boundary conditions the number of
defect lines corresponding to polymers surrounding the cycles is
always even. The situation for open boundary conditions is shown
in the right-hand figure. The branches of the spanning web going
to the root can be separated from the rest of the web by an
additional polymer line (marked by 5 in Fig. \ref{densed}). Thus,
the total number of defect lines becomes odd and one obtains the
set of even entries of the Kac table.

\section{Conformal field theory of spanning webs}

The partition function (\ref{conf-ZNM}) has interpretation in terms of
symplectic fermions \cite{Kausch}.
The symplectic fermions are fermionic fields $\theta^{\pm}(z)$ with
operator product expansion
\begin{equation}
\theta^{+}(z)\theta^{-}(w)\sim \log(z-w).
 \label{ProdExpFermions}
\end{equation}
These fields admit periodic and antiperiodic boundary conditions for
which they decompose with integer $\theta^{\pm}_n$,
$n \in \mathbb{Z}$,
and half-integer $\theta^{\pm}_n$, $n \in
\mathbb{Z}+\frac{1}{2}$, modes,
respectively. These modes satisfy the anticommutation relations
\begin{equation}
[\theta^{+}_n,\theta^{-}_m]_{+}=n\delta_{n+m,0}.
 \label{CommutatorFermionsPeriodic}
\end{equation}
Let $\mathcal{A}(2)$ denote this infinite dimensional Clifford
algebra. Strictly speaking, $\mathcal{A}(2)$ is not an algebra
or a vertex-operator algebra because
multiplication between integer and half integer modes is not defined,
but $\mathcal{A}(2)$ is very similar to ordinary vertex-operator algebras
and many standard notions can be defined for it (see discussion on this
subject in \cite{FT}).

The algebra $\mathcal{A}(2)$ contains the Virasoro subalgebra
generated by the energy-momentum tensor
\begin{equation}
T(z)=:\partial\theta^{+}(z)\partial\theta^{-}(z):
 \label{TensorEnergy}
\end{equation}
with central charge $c=-2$.

The algebra $\mathcal{A}(2)$ has two irreducible modules $X_1$ and
$X_2$ (see details in \cite{FFT,FT}). Modules $X_1$ and $X_2$ are
cyclic with cyclic (vacuum) vectors $|11\rangle$ satisfying
$\theta^{\pm}_n|11\rangle=0$ for $n\geq0$ and $|01\rangle$
satisfying $\theta^{\pm}_n|01\rangle=0$ for $n\geq\frac{1}{2}$
respectively. Module $X_1$ is generated by integer modes
$\theta^{\pm}_n$ with $n\leq -1$ from the vacuum vector
$|11\rangle$. Module $X_2$ is generated by half-integer modes
$\theta^{\pm}_n$ with $n\leq -\frac{1}{2}$ from the vacuum vector
$|01\rangle$. The algebra $\mathcal{A}(2)$ has two projective
modules $P_1$ and $P_2=X_2$. The module $P_1$ contains 4
irreducible subquotients isomorphic to $X_1$ \cite{FT}. The module
$P_1$ is cyclic with vacuum vector $|00\rangle$ satisfying
$\theta^{\pm}_n|00\rangle=0$ for $n\geq1$ and is generated from
$|00\rangle$ by integer modes $\theta^{\pm}_n$ with $n\leq 0$.

For a $\mathcal{A}(2)$-module $X$, we define the character
\begin{equation}
\chi(q,z)=\mathrm{Tr}_X q^{L_0-\frac{c}{24}}\omega^h
 \label{Character}
\end{equation}
where $L_0=\frac{1}{2\pi i }\oint z T(z)dz$ and $h$ is the operator
calculating the difference between the numbers of $\theta^{+}$ and $\theta^{-}$
modes in a state.

The character of $X_1$ is
\begin{equation}
\chi^{(1,1)}(q,\omega)=\frac{q^{\frac{1}{12}}}{\prod_{n=1}^{\infty} (1-q^n )}
  \sum_{r \in \mathbb{N}}
\sum_{j=0}^{r}\omega^{r-2j}q^{\frac{r(r-1)}{2}}(1-q^r).
 \label{CharacterX1}
\end{equation}
The character of $X_2$ is
\begin{equation}
\chi^{(0,1)}(q,\omega)=\frac{q^{-\frac{1}{24}}}{\prod_{n=1}^{\infty} (1-q^n )}
\sum_{r \in \mathbb{N}}
\sum_{j=0}^{r}\omega^{r-2j}q^{\frac{(r-1)^2}{2}}(1-q^{2r}).
 \label{CharacterX2}
\end{equation}
The character of $P_1$ is
\begin{equation}
\chi^{(0,0)}(q,\omega)=( 2+\omega+\omega^{-1})\chi^{(1,1)}(q,\omega).
 \label{CharacterP}
\end{equation}
The same characters can be written in terms of Virasoro characters
in the form
\begin{equation}
\chi^{(1,1)}(q,\omega)=\sum_{r \in \mathbb{N}}
\sum_{j=0}^{r}\omega^{r-2j} \chi_{r1}(q)
 \label{CharacterPVirasoro}
\end{equation}
and
\begin{equation}
\chi^{(0,1)}(q,\omega)= \sum_{r \in \mathbb{N}}
\sum_{j=0}^{r}\omega^{r-2j} \chi_{r2}(q)
 \label{CharacterX2Virasoro}
\end{equation}
where
\begin{equation}
\chi_{rs}(q)=
\frac{q^{\Delta_{rs}-\frac{c}{24}}(1-q^{rs})}{\prod_{n=1}^{\infty}(1-q^n)}
 \label{ChiRS}
\end{equation}
are Rocha-Caridi characters of the irreducible Virasoro
representations with conformal dimensions
\begin{equation}
\Delta_{rs}=\frac{(2r-s)^2 -1}{8}
 \label{DeltaRS}
\end{equation}
and $c=-2$.

Now we intend to identify (\ref{conf-ZNM}) with characters of some
coinvariants calculated in $X_1$, $X_2$ and $P_1$. (By
definition, the coinvariant in the $A$-module $X$ with respect to
the subalgebra $B\subset A$ is the quotient $X/BX$). We fix the
subalgebra $\mathcal{A}(2)[N]$ of $\mathcal{A}(2)$ for $N \in
\mathbb{N}$
\begin{equation}
\mathcal{A}(2)[N] = \left\{ \begin{array}{rll}
\{\theta^{\pm}_{-n}, \  n\geq N \}& \mathrm{periodic} \\
\{\theta^{\pm}_{-n-\frac{1}{2}}, \ n\geq N  \}&
\mathrm{antiperiodic}\end{array}      \right.
 \label{A2N}
\end{equation}
and consider the characters $\chi^{(1,1)}[N](q,\omega)$,
$\chi^{(0,1)}[N](q,\omega)$ and $\chi^{(0,0)}[N](q,\omega)$ of
coinvariants with respect to $\mathcal{A}(2)[N]$ in the modules
$X_1$, $X_2$ and $P_1$, respectively. These characters coincide
with (\ref{conf-ZNM})
\begin{equation}\label{identification}
 \chi^{(\mu,\nu)}[N](q,\omega) = \bar{Z}_{N}^{(\mu,\nu)}(q,\omega).
  \label{Chi}
 \end{equation}
For $N\rightarrow\infty$ one has
\begin{equation}
\chi^{(\mu,\nu)}[N](q,\omega) \rightarrow \chi^{(\mu,\nu)}(q,\omega).
 \label{ChiAsympt}
\end{equation}

The polynomials $C^{(1,1)}_{3,s}[N](q)$ and
$C^{(0,1)}_{3,s}[N](q)$ converge to the characters of irreducible
Virasoro modules
\begin{equation}
 C^{(1,1)}_{3,s}[N](q)\to\chi_{s1}(q),\quad
C^{(0,1)}_{3,s}[N](q)\to\chi_{s2}(q),
\end{equation}
and $C^{(0,0)}_{3,s}[N](q)$ converges to the characters of
logarithmic Virasoro modules for the $\mathcal{LM}(1,2)$ model
\cite{PRZ} as $N$ tends to infinity. Note also that characters
(\ref{Chi}) can be expressed through the Kostka-like polynomials
$\hat{K}^{(2)}_{\ell,N}(q,\omega)$ from \cite{FT} in the following
way
\begin{equation}
 \chi^{(1,1)}[N](q,\omega)=q^{\frac{1}{12}}\hat{K}^{(2)}_{1,2N}(q,\omega),\qquad
  \chi^{(0,1)}[N](q,\omega)=q^{-\frac{1}{24}}\hat{K}^{(2)}_{2,2N+1}(q,\omega).
\end{equation}

Now we can identify open and closed boundary conditions of the
spanning webs model with irreducible modules of the algebra
$\mathcal{A}(2)$. In conformal field theory, the boundary
conditions are in one to one correspondence with irreducible
modules of the chiral algebra \cite{BPPZ}. For the partition
function $\bar{Z}^{(\mu,\nu)}$ on a cylinder with boundary
conditions $\mu$ and $\nu$ the conformal field theory predicts
\begin{equation}
 \bar{Z}^{(\mu,\nu)}(q)=\sum_{\gamma} N_{\mu\nu}^\gamma \chi_\gamma(q),
\end{equation}
where $N_{\mu\nu}^\gamma$ are structure constants in the fusion
between modules labelled by $\mu$ and $\nu$ and $\chi_\gamma(q)$
are characters of modules appearing in this fusion. The fusion of
$\mathcal{A}(2)$ modules is
\begin{equation}
 X_1\dot\otimes X_1=X_1,\quad X_1\dot\otimes X_2=X_2,\quad X_2\dot\otimes X_2=P_1.
\end{equation}
The fusion together with identification (\ref{identification}) leads to the correspondence
\begin{equation}
 \mbox{open b.c.}\leftrightarrow X_1\qquad \mbox{closed b.c.}\leftrightarrow X_2.
\end{equation}

An identification of triplet $W$  algebra or Virasoro boundary
conditions \cite{PRR} is more subtle because of their nonlocal
nature. We start the consideration with $W$ boundary conditions.
Each $\mathcal{A}(2)$-module decomposes into direct sum of two
$W$-modules, which are labelled by the same symbol with additional
superscript $\pm$
\begin{equation}
 X_1=X_1^+\oplus X_1^-,\quad X_2=X_2^+\oplus X_2^-,\quad P_1=P_1^+\oplus P_1^-.
\end{equation}
The characters of $W$ modules are obtained from characters of $\mathcal{A}(2)$ modules
by taking odd and even parts in $\omega$.
For example
\begin{equation}
 \chi_1^\pm(q)=\frac{1}{2}\left[\chi^{(1,1)}(q,1)\pm\chi^{(1,1)}(q,-1)\right],
\end{equation}
where $\chi_1^\pm(q)$ are the characters of $X^\pm_1$. We can
interpret this in lattice terms as taking only configurations with
even or odd differences between numbers of non-contractible loops
of different orientations (see also the paragraph after Eq.
(\ref{C polynom3})). However, it is not clear how to formulate
such conditions as local boundary conditions without reference to
the bulk.

In order to establish a connection with boundary conditions corresponding to
Virasoro representations we note that the algebra $\mathcal{A}(2)$ admits
a $sl(2)$ action such that $\theta^+(z)$ and $\theta^-(z)$ are the highest and
lowest weight vectors of the doublet. This $sl(2)$ action commutes with
the Virasoro subalgebra (\ref{TensorEnergy}) and therefore irreducible
$\mathcal{A}(2)$ modules decompose as
\begin{equation}
 X_{1}=\oplus_{r\in\mathbb{N}}\pi_r\otimes Y_{r1},\quad
  X_{2}=\oplus_{r\in\mathbb{N}}\pi_r\otimes Y_{r2},
\end{equation}
where $\pi_r$ is $r$ dimensional irreducible $sl(2)$
representation and $Y_{rs}$ is the Virasoro irreducible
representation with the conformal dimension (\ref{DeltaRS}). The
character of $\pi_r$ is $\sum_{j=0}^{r}w^{r-2j}$, which explains
the decompositions (\ref{CharacterPVirasoro}),
(\ref{CharacterX2Virasoro}) and our definition of polynomials
$C^{(\mu,\nu)}_{3,s}[N](q)$ in (\ref{N polynom3}). A lattice
interpretation of $C^{(\mu,\nu)}_{3,s}[N](q)$ is very cumbersome
and appeals to conditions on non-contractible loops in the bulk
like in the $W$ case.

\section*{Conclusions}

In this paper we have found the exact partition function for a model of spanning webs
parameterized by the number of non-contractible cycles for the finite cylinder geometry.
We have calculated the leading finite-size corrections and identified them with the finitized
characters for the minimal logarithmic conformal field theory with $c=-2$.

The model considered here is similar in many aspects to the model of critical dense polymers
solved by Pearce and Rasmussen \cite{PearceRas} by using the planar Temperley-Lieb algebra
and commuting double-row transfer matrices. There are, however, several features which are
different in these models. First, the cylinder geometry admits classification of the webs
configurations in terms of numbers of non-contractible cycles which are well defined
"quantum numbers". Regarding the conservation law in the system of dense
polymers, one notices that the transfer matrices used in \cite{PearceRas} have a
block triangular structure. This structure reflects the fact that defect lines can be
annihilated in
pairs and, therefore, the number of defects is not conserved. Imbedding the
system into the cylinder geometry, which is the case of our model, is
equivalent to taking the trace of a transfer matrix, hence, the number of non-contractible
cycles (or defect lines in the
case of dense polymers) becomes automatically fixed. The simple geometry of the cylinder with
closed or open boundary conditions on the edges allows an elementary evaluation
of the partition function by using an extension of the Kirchhoff theorem.

The second and more important feature of the spanning webs model consists in the perfect
coincidence of the universal part of its partition function $\bar{Z}_{N}^{(\mu,\nu)}(q,\omega)$
for different combinations of closed and open boundary conditions, $\mu =0,1$ and
$\nu =0,1$, with the finitized characters of the
symplectic fermions, see (\ref{Chi}). This allows us to interpret the simplectic fermion model as a
conformal field theory
of spanning webs on a cylinder. The further identification of the triplet $W$ algebra in terms of
spanning webs is an interesting open problem. 

Another problem for future investigation is the
explicit construction of Virasoro representations with finitized characters given by
Eqs. (36) - (38). Strictly speaking we have not proved the indecomposability of these
Virasoro representations. However, the structure of characters (36) - (38) gives indication
that our model belongs to the universality class of the $c=-2$ logarithmic conformal field
theory (LCFT). An additional argument in
support of the above conjecture provides the one-to-one correspondence between the recurrent
configurations of the Abelian sandpile model (ASM) and the spanning trees on the square lattice
\cite{MD}. Under this mapping the height variables of the ASM correspond to nonlocal 
correlations in the spanning trees. Explicit calculations of height correlations in the ASM
show logarithmic corrections \cite{PR05} in complete agreement with the predictions 
of the $c=-2$ LCFT.

\section*{Acknowledgments}
We thank A. Gainutdinov, P. Pearce, V. Rittenberg and P. Ruelle for helpful comments and discussions.
This work was supported by RFBR grant No 06-01-00191a and a JINR - Bulgaria
collaboration grant. The work of IYuT was supported in part by LSS-1615.2008.2,
the RFBR Grant 08-02-01118 and the ``Dynasty'' foundation.

\end{document}